\documentclass[aip,twocolumn,amsmath,amssymb,seceqn,10pt]{revtex4-1}
\usepackage{graphicx}
\usepackage{graphics}
\usepackage{epsfig}
\usepackage{dcolumn}
\usepackage{bm}
\usepackage{makeidx}
\usepackage{subfigure}
\usepackage{color}
\usepackage{longtable}

\begin{document}

\title{Antiferromagnetic spintronics}
\author{T. Jungwirth}
\affiliation{Institute of Physics ASCR, v.v.i., Cukrovarnick\'a 10, 162 53 Praha 6, Czech Republic}
\affiliation{School of Physics and Astronomy, University of Nottingham, Nottingham NG7 2RD, United Kingdom}
\author{X.~Marti}
\affiliation{Institute of Physics ASCR, v.v.i., Cukrovarnick\'a 10, 162 53 Praha 6, Czech Republic}
\author{P.~Wadley}
\affiliation{School of Physics and Astronomy, University of Nottingham, Nottingham NG7 2RD, United Kingdom}
\author{J. Wunderlich}
\affiliation{Institute of Physics ASCR, v.v.i., Cukrovarnick\'a 10, 162 53 Praha 6, Czech Republic}
\affiliation{Hitachi Cambridge Laboratory, Cambridge CB3 0HE,  United Kingdom}


\begin{abstract}
Antiferromagnetic materials are magnetic inside, however, the direction of their ordered microscopic moments alternates between individual atomic sites. The resulting zero net magnetic moment  makes magnetism in antiferromagnets invisible on the outside. It also implies that if information was stored in antiferromagnetic moments it would be insensitive to disturbing external magnetic fields, and the antiferromagnetic element would not affect magnetically its neighbors no matter how densely the elements were arranged in a device. The intrinsic high frequencies of antiferromagnetic dynamics represent another property that makes antiferromagnets distinct from ferromagnets. The outstanding question is how to efficiently manipulate and detect the magnetic state of an antiferromagnet. In this article we give an overview of recent works addressing this question. We also review studies looking at merits of antiferromagnetic spintronics from a more general perspective of spin-ransport, magnetization dynamics, and materials research, and give a brief outlook of future research and applications of antiferromagnetic spintronics.
\end{abstract}

\maketitle

Interesting and useless - this was the common perception of antiferromagnets expressed quite explicitly, for example, in the 1970 Nobel lecture of Louis N\'eel.\cite{Neel1970} Connecting to this traditional notion we can define antiferromagnetic spintronics as a field that makes antiferromagnets useful and spintronics more interesting. Below we give an overview of this emerging field  whose aim is to complement or replace ferromagnets in active components of spintronic devices. 

We  recall some key physics roots of the field and first concepts of spintronic devices based on antiferromagnetic counterparts of the non-relativistic giant-magnetoresistance and spin-transfer-torque phenomena.\cite{MacDonald2011} We then focus on electrical reading and writing of information, combined with robust storage, that can be realized in antiferromagnetic memories via relativistic magnetoresistance and spin torque effects.\cite{Shick2010,Zelezny2014} Related to these topics is the research of spintronic devices in which antiferromagnets act as efficient generators, detectors, and transmitters of spin currents. This will lead us to studies exploring fast dynamics in antiferromagnets\cite{Gomonay2014} and different types of antiferromagnetic materials. They range from insulators to superconductors. Here we comment also on the relation between crystal antiferromagents and synthetic antiferromagnets, with the latter ones playing an important role in spintronic sensor and memory devices.\cite{Parkin2003} In concluding remarks we outline some of the envisaged future directions of research and potential applications of antiferromagnetic spintronics.

\subsection*{Equilibrium properties and magnetic storage in antiferromagnets}

The understanding of equilibrium properties of ferromagnets has been guided by the notion of a {\em global molecular field}, introduced by Pierre Weiss.\cite{Neel1970} The theory starts from the Curie law for paramagnets with the inverse susceptibility proportional to temperature, $\chi^{-1}\sim T$. It further assumes that the externally applied uniform magnetic field is accompanied in ferromagnets by a uniform internal molecular field, $\lambda M$, proportional to the magnetization $M$ and the Weiss molecular field constant $\lambda$. The high-temperature inverse susceptibility of ferromagnets is then described by the Curie-Weiss law, $\chi^{-1}\sim T-\theta$, where $\theta>0$ is the Curie constant proportional to $\lambda$. The microscopic origin of the molecular field was explained by Heisenberg in terms of the exchange interaction between neighboring magnetic atoms favoring parallel alignment of their magnetic moments and leading to the ferromagnetic order with a large macroscopic moment below the Curie temperature. 

In early 1930's, Loius N\'eel was drawn into the problem that some materials containing magnetic elements and showing zero remanence at all temperatures did not follow the paramagnetic Curie law.\cite{Neel1970} Instead they  obeyed the Curie-Weiss law at high temperatures, however with a negative $\theta$, and showed a nearly constant susceptibility at low temperatures. Since at high temperatures the magnetic atoms with strongly thermally fluctuating  moments can be considered identical, the global molecular field could still be invoked, albeit with a negative $\lambda$ to explain the negative Curie constant. N\'eel pointed out that the microscopic origin of the negative Weiss molecular field is in the exchange interaction between neighboring magnetic atoms favoring anti-parallel alignment of their moments. He emphasized that this interaction is not compatible with a low-temperature ordered state that can be described by a global uniform molecular field. Instead he introduced the concept of a {\em local molecular field} which can vary  at inter-atomic length scales.\cite{Neel1970} 

Using an example of two interlaced cubic sublattices, N\'eel described a new type of magnetic order in which the local molecular field had opposite sign on the two sublattices, stabilizing a spontaneous magnetization of one sign on the first sublattice and of the opposite sign on the second sublattice. In magnetically isotropic systems, i.e. when neglecting the relativistic coupling between spins and the lattice, an infinitesimally weak external magnetic field would align the antiparallel sublattice magnetizations along an axis parpendicular to the applied field. With increasing field strength, the magnetic sublattices increasingly tend to cant their moments towards the field. This leads to the development of a non-zero net moment whose amplitude is inverse proportional to the local molecular field constant (to the exchange coupling between the sublattices), proportional to the external magnetic field, and independent of temperature. This was N\'eel's explanation of the constant low-temperature susceptibility seen, e.g., in the elemental metal of Cr and later in a number of systems called antiferromagnets.\cite{Neel1970}

Apart from introducing  the concept of the local molecular field, several other observations made in N\'eel's seminal works have provided key principles for the development of antiferromagnetic spintronics. N\'eel noted a general rule that  {\em effects depending on the square of the spontaneous (sublattice) magnetization should show the same variation in antiferromagnets as in ferromagnets}.\cite{Neel1970} One example he considered was the magnetic anisotropy energy. In ferromagnetic memories,\cite{Chappert2007} it is this quantity that provides the energy barrier separating two different stable directions of ordered spins, representing 1 and 0. Storing magnetic information in devices made of antiferromagnets should, therefore, be equally feasible, as confirmed in several recent experiments.\cite{Park2011b,Loth2012,Marti2014,Wadley2015,Moriyama2015} 

\begin{figure}[h!]
\hspace*{-0cm}\epsfig{width=1\columnwidth,angle=0,file=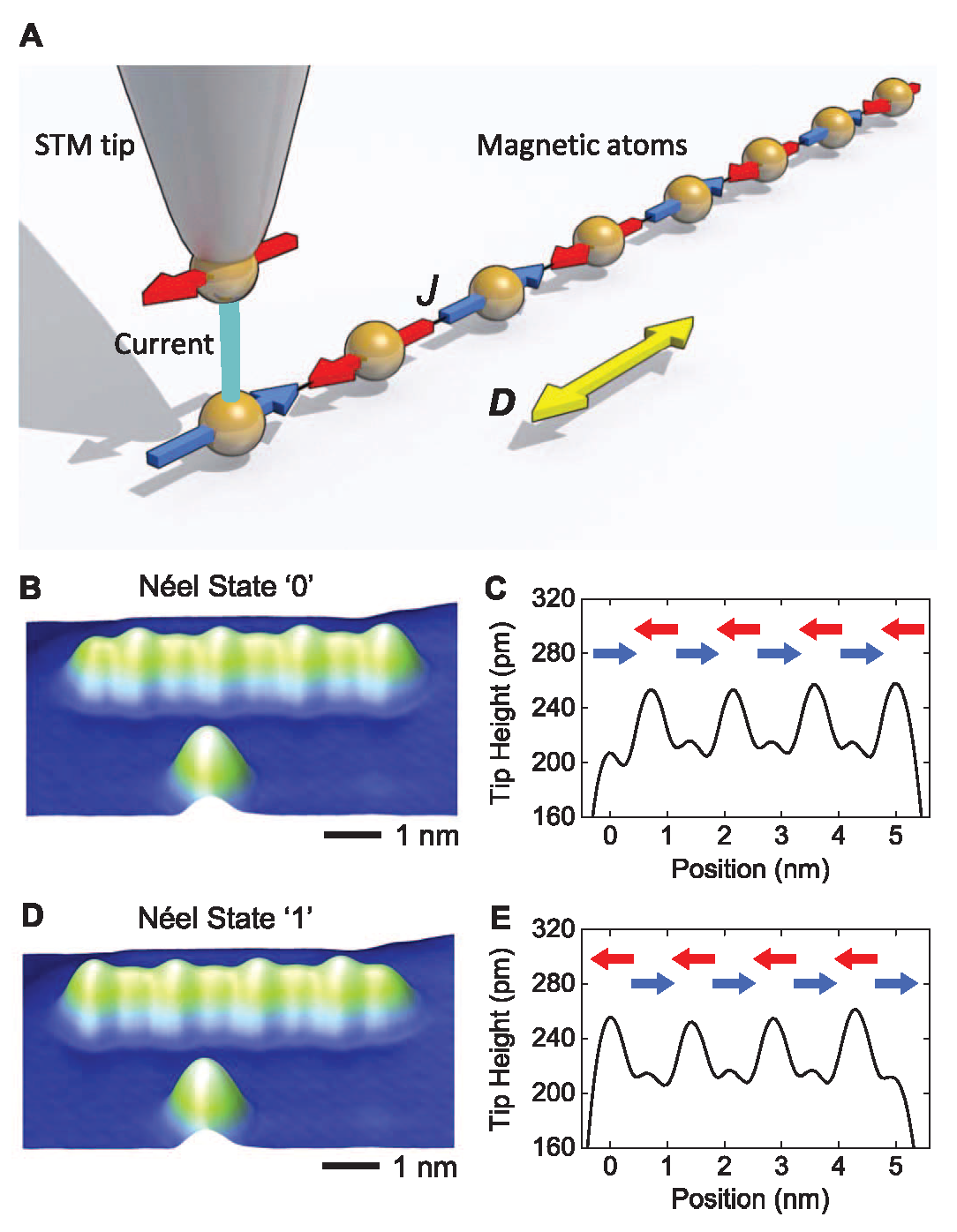}
\caption{{\bf a}, Schematic of atoms on a surface coupled antiferromagnetically with exchange energy $J$. Surface-induced magnetic anisotropy fields cause the spins of the atoms to align parallel to the easy magnetic axis, $D$. A spin-polarized STM tip reads the magnetic state of the structure by magnetoresistive tunneling, atom-by-atom. A magnetic field applied parallel to $D$ polarizes the tip. {\bf b}, Spin-polarized STM image of a linear chain of eight Fe atoms. Spins are in N\'eel state 0. {\bf c}, Section through center of chain in {\bf b}, with the spin orientation of each Fe atom indicated by colored arrows. ({\bf d}, and {\bf e},) same as {\bf b}, and {\bf c}, but in N\'eel state 1. From Ref.~\onlinecite{Loth2012}.}
\label{Fe_chain}
\end{figure}

Fig.~\ref{Fe_chain} shows an example of storing information, at temperatures of a few Kelvin, in a nanostructure comprising an antiferromagnetic chain of Fe atoms.\cite{Loth2012} A polarized scanning tunneling microscope tip sets and detects, atom-by-atom, two distinct stable states of the antiferromagnetically coupled Fe spins. The measurements highlight current experimental capabilities, unthinkable at times of N\'eel's seminal works, of the control of antiferromagnetic moments by  aiming the external probe at a specific individual atom, belonging to one or the other magnetic sublattice. The experiment provides a direct microscopic image of information storage in an antiferromagnet. Moreover, it vizualizes with the ultimate atomistic resolution the N\'eel's local field principle extended to non-equilibrium phenomena for writing and reading information in antiferromagnets. On the other hand, controlling a few spin nanostructure by spin polarized STM resembles a mechanical hard-drive taken in a laboratory environment to the ultimate atomistic limit. This scheme does not open a route to antiferromagnetic spintronics compatible with practical approaches for designing microelectronic devices. In the following section we show that in antiferromagnets with many magnetic atoms one can find physical phenomena that allow for the seemingly impossible local control of the antiferromagnetic spin-sublattices by global electrical currents  in common microelectronics set-ups.  Simultaneously, the sufficiently large number of magnetic atoms in these devices provides robust storage at room temperature. 

Before that, still in the context of this section focusing on equilibrium properties, we recall one more aspect of N\'eel's pioneering studies. When analyzing the magnetic susceptibility in the presence of the magneto-crystalline anisotropy, N\'eel concluded that for the magnetic field applied along the antiferromagnetic easy axis the susceptibility is zero up to a spin-flop field that scales with the geometric mean of the exchange and anisotropy fields. Above the spin-flop field, the antiferromagnetic moments switch to a direction perpendicular to the field, resulting in the above constant susceptibility inverse-proportional to the exchange energy. In ferromagnets, on the other hand, magnetization is reoriented by magnetic fields proportional to the anisotropy fields. Relativistic or dipolar magnetic anisotropy fields are many orders of magnitude weaker in typical magnets than exchange fields. Antiferromagnets therefore not only generate zero stray fields and by this automatically eliminate unintentional magnetic cross-talk between neighboring devices, but also provide magnetic storage that is exceptionally robust against magnetic field perturbations.\cite{Marti2014}

\subsection*{Writing and reading magnetic state in antiferromagnets}
The insensitivity to magnetic fields comes at a price of the notorious difficulty of manipulating antiferromagnetic moments by means comparably efficient to ferromagnets.
One possibility is offered by the exchange-coupling at an interface between an antiferromagnet and a ferromagnet.\cite{Nogues1999,Scholl2004} The effect is already utilized in ferromagnetic spin-valves, comprising a pair of fixed and free ferromagnetic layers and forming the basis of commercial magnetic field sensors and magnetic random access memories (MRAMs).\cite{Chappert2007} Exchange-coupling to an antiferromagnet enhances the magnetic hardness of the fixed reference layer.\cite{Nogues1999} In this arrangement, the antiferromagnetic moments are assumed to be also fixed and the antiferromagnet plays only a passive supporting role in the spintronic device. In another arrangement where the ferromagnet is soft and the adjacent antiferromagnet thin enough, a weak external magnetic field can reorient the ferromagnet whose interfacial moments then drag the neighboring antiferromagnetic moments via the interfacial exchange spring. The method was already employed to control antiferromagnetic moments by weak fields in several studies of spintronic devices.\cite{Park2011b,Duine2011,Wang2012a,Ralph2013,Fina2014,Wang2014e}

\noindent{\bf\em Electrical control by non-relativistic effects.} In MRAMs, the trend is to abandon writing by magnetic fields because the method is not scalable.\cite{Chappert2007} The most extensively explored alternative is writing by the current-induced spin transfer torque (STT). It is basically a non-relativistic phenomenon understood in terms of the global angular momentum conservation and the corresponding {\em transfer from the carrier spin angular momentum to the magnetization angular momentum}.\cite{Ralph2008} 

The STT is considered to be driven by an effective field proportional to the non-equilibrium carrier spin polarization ${\bf s}$ in the free recording ferromagnet and to have the general form ${\bf T}=d{\bf M}/dt\sim {\bf M}\times{\bf s}$, where ${\bf M}$ is the magnetization in the free ferromagnet. In the limit of a short carrier spin life-time relative to the spin precession time in the free ferromagnet, ${\bf s}\sim{\bf p}$ is independent of ${\bf M}$ and proportional to the polarization ${\bf p}$ of the injection spin current defined by the fixed ferromagnetic polarizer. Magnetization dynamics induced by this field-like STT, ${\bf T}\sim {\bf M}\times{\bf p}$, is analogous to applying an external magnetic field. Switching then occurs when the current-induced effective field overcomes the magnetic anisotropy fields in the free ferromagnet.

In the limit of long carrier spin life-time, the injected carrier spins precess around the magnetization of the free ferromagnet. The resulting ${\bf s}\sim{\bf M}\times{\bf p}$ depends on ${\bf M}$ in this case. The corresponding (anti)damping-like STT, ${\bf T}\sim {\bf M}\times({\bf M}\times{\bf p})$,  can contribute or compete with the Gilbert damping in the ferromagnet, depending on the polarity of the applied current between the fixed and the free ferromagnet. In different configurations, switching by this type of torque occurs when the current-induced effective field overcomes the anisotropy fields, or anisotropy fields rescaled by the magnetization damping factor which is typically $\ll 1$. In common transition metal ferromagnets, relatively long carrier spin life-times imply that the (anti)damping-like STT typically dominates. Electrical switching by the STT in ferromagnetic spin valves is reversible by flipping the sign of the writing current.
\begin{figure}[h!]
\hspace*{-0cm}\epsfig{width=1\columnwidth,angle=0,file=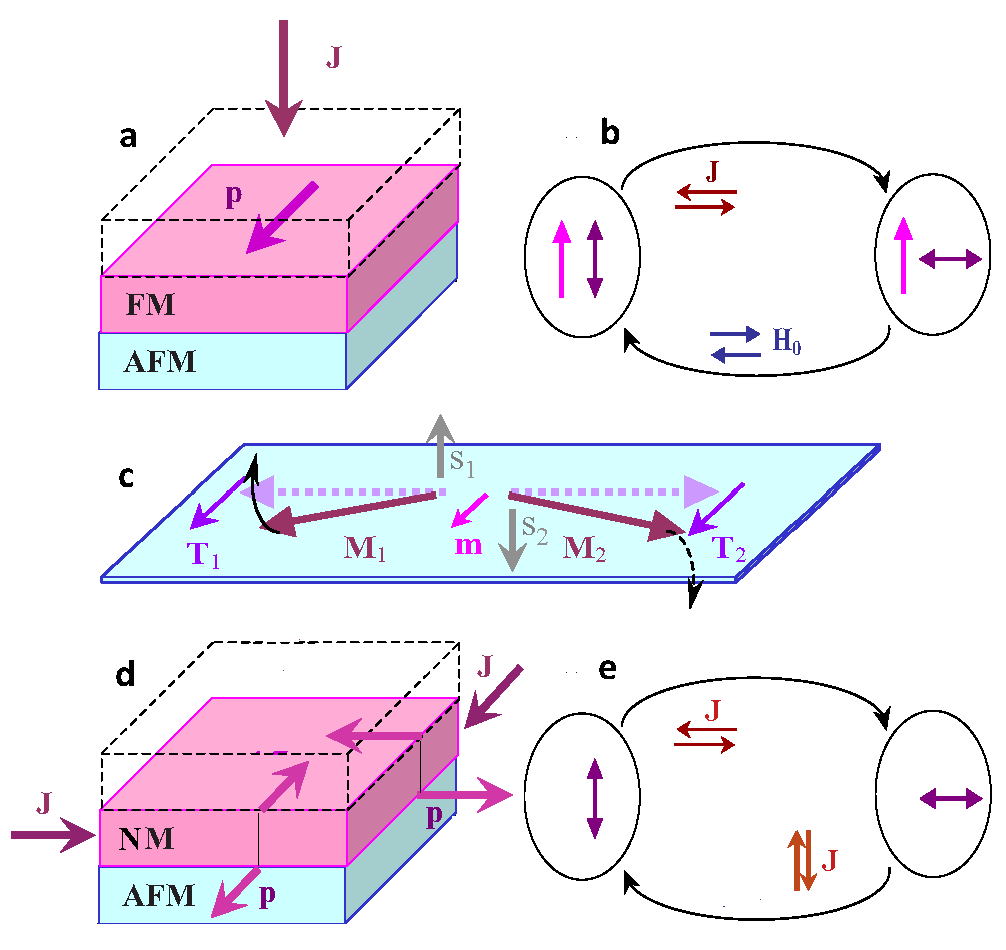}
\caption{{\bf a},  
Schematic of a ferromagnet/antiferromagnet bilayer. ${\bf J}$ is the vertical electrical current  and ${\bf p}$ is the polarization of electrons injected from the fixed ferromagnet. {\bf b}, Transition from parallel to perpendicular configuration  of the ferromagnet and antiferromagnet spin axes can be induced by current of either polarity. Transition from perpendicular to parallel configuration can be induced only by a spin-flop magnetic field, ${\bf H}_0$, assuming a fixed ferromagnet. {\bf c}, Local (anti)damping-like STTs, ${\bf T}_1$ and ${\bf T}_2$,  driven by staggered local non-equilibrium carrier polarization, ${\bf s}_1\sim {\bf M}_1\times{\bf p}$ and ${\bf s}_2\sim {\bf M}_2\times{\bf p}$, on antiferromagnetic sublattices with magnetizations ${\bf M}_1$ and ${\bf M}_2$ shown before (dotted arrows) and after (solid arrows) the action of the STT (${\bf m}$ is the canting moment).  
From Ref.~\onlinecite{Gomonay2010}. {\bf d},  Vertical injection of a spin-current in a non-magnet/antiferromagnet bilayer due to a lateral electrical current and SHE in the non-magnetic layer. {\bf e}, Reversible transition between the two configurations of the antiferromagnet can be controlled by orthogonal in-plane electrical currents. From Ref.~\onlinecite{Zelezny2014}.}
\label{F-AF}
\end{figure}

For antiferromagnets, the STT phenomenology is modified  by considering a spin current into a particular atomic site that tends to produce a torque which acts on the spin centred on that site, attempting to restore spin conservation locally.\cite{MacDonald2011} For a given spin sublattice $i$ of the antiferromagnet, the local STT is given by the local sublattice magnetization and the local non-equilibrium carrier spin polarization, ${\bf T}_{i}\sim {\bf M}_{i}\times{\bf s}_i$. As in ferromagnets, two types of  local non-equilibrium carrier spin polarizations can be considered, ${\bf s}_i\sim {\bf p}_i$ and  ${\bf s}_i\sim {\bf M}_i\times{\bf p}_i$, corresponding to the local field-like STT, ${\bf T}_{i}\sim {\bf M}_{i}\times{\bf p}_i$, and (anti)damping-like STT, ${\bf T}_{i}\sim {\bf M}_{i}\times({\bf M}_{i}\times{\bf p}_i)$, respectively. 

Unlike the STM experiment\cite{Loth2012} discussed in the previous section, here the manipulation of the antiferromagnet by the local non-equilibrium effective fields is considered to be driven by a  global uniform electrical current. {\em The local control by global currents is the key prerequisite for making antiferromagnetic microelectronics feasible.}

Figs.~\ref{F-AF}a-c show conceptually an example of the (anti)damping-like STT in a fixed-ferromagnet/antiferromagnet bilayer excited by a vertical electrical current.\cite{Gomonay2010} Here the injected spin polarization form the reference ferromagnet  is the same for both spin sublattices in the antiferromagnet, i.e.  ${\bf p}_1={\bf p}_2={\bf p}$. It implies that the local non-equilibrium spin polarizations, ${\bf s}_1\sim {\bf M}_1\times{\bf p}$ and ${\bf s}_2\sim {\bf M}_2\times{\bf p}$, have opposite sign on the two spin sublattices since ${\bf M}_1=-{\bf M}_2$. The corresponding non-equilibrium field $\sim{\bf s}_i$ is, therefore, also staggered. This makes it equally efficient in the antiferromagnet as  uniform current-induced fields that generate (anti)damping-like STTs in ferromagnets. For a uniform injection polarization ${\bf p}$, the (anti)damping-like STT is an even function of the global magnetization in ferromagnets or local spin-sublattice magnetization in antiferromagnets. The comparable efficiency in both types of magnetic systems reminds us again of the general N\'eel's principle of the similarity between ferromagnets and antiferromagnets in $M^2$-dependent quantities.

On the other hand, the field-like STT in the antferromagnet would be driven in the geometry of Fig.~\ref{F-AF}a by a uniform non-staggered effective field $\sim{\bf p}$, i.e., would be equally inefficient as a uniform external magnetic field acting on an antiferromagnet. We summarize that large reorientations of the antiferromagnetic moments by weak effective current-induced fields, comparable to the anisotropy fields (possibly reduced by the damping factor), require staggered local effective fields, i.e., uniform non-staggered local torques (see Fig.~\ref{F-AF}c).

The efficient (anti)damping-like STT in the geometry of Fig.~\ref{F-AF}a can induce a switching from a parallel to a perpendicular configuration of the antiferromagnetic moments with respect to the fixed ferromagnet, as shown in Fig.~\ref{F-AF}b. This is, however, independent of the polarity of the vertical electrical current so the antiferromagnet cannot be electrically switched back to the parallel configuration. A large spin-flop magnetic field has to be applied to reverse the state.\cite{Gomonay2010} Moreover, the structure comprises the auxiliary reference ferromagnet which diminishes some of the merits of spintronics based on antiferromagnets alone.

When using an antiferromagnet instead of the ferromagnet as the reference spin injector, the polarization ${\bf p}_i$ of the transmitted  electrons through the reference antiferromagnet can oscillate with a period commensurate with its antiferromagnetic order.\cite{Nunez2006,Saidaoui2014} By adding to the structure a second, free antiferromagnet with a commensurate lattice one can infer from the above considerations the symmtries of the STTs acting in the second antiferromagnet. Since ${\bf p}_1=-{\bf p}_2$ is staggered in this case, the effective field $\sim{\bf s}_i\sim {\bf M}_i\times{\bf p}_i$, driving the (anti)damping-like STT, is non-staggered and is therefore inefficient. In this case the efficient torque is the field-like STT driven by a staggered, magnetization-independent effective field $\sim{\bf p}_i$. As mentioned above, the field-like STT tends to have the weaker amplitude of the two types of torques in common transition metals. Moreover, microscopic calculations showed that in these all-antiferromagnetic spin valves, the non-relativistic STTs are subtle, spin-coherent quantum-interference phenomena relying on perfectly epitaxial and commensurate multilayers.\cite{Nunez2006,MacDonald2011,Saidaoui2014} This may explain why the STT in antiferromagnetic spin valves has not yet been identified experimentally.

Disorder is also detrimental to the reading scheme proposed for the antiferromagnetic spin valves within the framework of non-relativistic  spintronics.\cite{Nunez2006,MacDonald2011} The proposal refers to the giant/tunneling magnetoresistance (GMR/TMR) in ferromagnetic spin valves with conductive/insulating non-magnetic spacer whose resistance depends on the relative orientation of the magnetization in the reference and free ferromagnet.\cite{Chappert2007} In antiferromagnetic spin-valves with perfectly epitaxial commensurate multilayers, it is the relative orientation of the local spins on the last atomic planes of the two antiferromagnets facing each other across the non-magnetic spacer that determines  the read-out resistance signal.\cite{Nunez2006} The difficulty to observe the effect experimentally has casted doubts on the principle ability to detect by practical means any effects of current on the magnetic order of an antiferromagnet.\cite{MacDonald2011}  The attention within the non-relativistic spintronics framework thus turned back to interfaces of antiferromagnets with ferromagnets\cite{Wei2007,Gomonay2010,Gomonay2012,Gomonay2014,Prakhya2014,Cheng2014,MacDonald2011} and to indirect observations of effects in the antiferromagnet by measuring induced magnetic signals in the adjacent exchange-coupled ferromagnet.\cite{Tang2007,Urazhdin2007,Dai2008,Tang2012,MacDonald2011} 

\noindent{\bf\em Electrical control by relativistic effects.} Relativistic physics provides the means for electrical read-out of the orientation of the antiferromagnetic moments in bulk antiferromagnets and interfaces.\cite{Shick2010} To comprehend this we can recall again N\'eel's principle of the correspondence between ferromagnets and antiferromagnets in $M^2$-dependent phenomena. We already mentioned that the relativistic magneto-crystalline anisotropy energy is one example that has this property. Its relativistic magneto-transport counterpart is the anisotropic magnetoresistance (AMR).\cite{McGuire1975} 

\begin{figure}[h!]
\hspace*{-0cm}\epsfig{width=1\columnwidth,angle=0,file=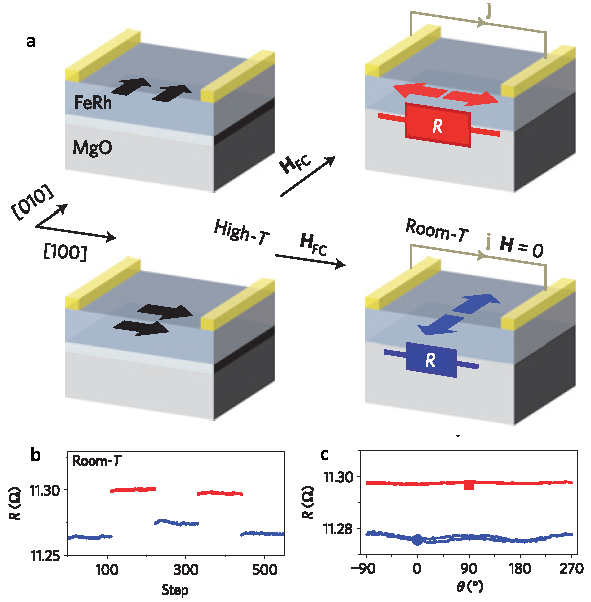}
\caption{{\bf a}, Schematic illustration of the FeRh memory. For writing, the sample is cooled in a field $H_{\rm FC}$ from a temperature above the antiferromagnetic-ferromagnetic transition in FeRh. Black arrows denote the orientation of the magnetic moments in the ferromagnetic phase whereas either red or blue arrows denote two distinct configurations of the magnetic moments in the antiferromagnetic phase. {\bf b}, Resistance measured at room temperature and zero magnetic field after field-cooling the sample with field  parallel (blue) and perpendicular (red) to the current direction.  {\bf c}, Stability of the two memory states at room temperature tested by measuring the resistance while rotating a 1 T magnetic field. The states cannot be erased by fields as high as 9~T. From Ref.~\onlinecite{Marti2014}.}
\label{FeRh}
\end{figure}

The first generation of spintronic magnetic field sensors and MRAMs made in ferromagnets used ohmic AMR.\cite{Daughton1992} Recently, several experiments have demonstrated  AMR read-out  in ohmic antiferromagnetic devices.\cite{Marti2014,Fina2014,Zhang2014b,Wong2014,Moriyama2015,Kriegner} Fig.~\ref{FeRh} shows an example of an FeRh antiferromagnetic memory resistor\cite{Marti2014} in which one state has antiferromagnetic moments aligned parallel and the other state perpendicular to the probing current direction. This allows to use the AMR for the detection. The read-out via a global electrical current is combined in this memory device with a room-temperature storage whose insensitivity to magnetic field perturbations at room temperature was tested and confirmed up to 9~T. For setting the two distinct states, FeRh was heated above the transition to a ferromagnetic state and then cooled back to the room temperature antiferromagnetic phase with a writing magnetic field applied along one of the two orthogonal directions.\cite{Marti2014} The heat-assisted magneto-recording in the antiferromagnetic FeRh memory can be realized using electrically generated Joule heating.\cite{Moriyama2015} 

AMR signals in ferromagnets and antiferromagnets are typically limited to a few per cent which, together with the low resistivity, makes  ohmic AMR devices unfavorable for high density MRAMs.\cite{Parkin2003} For this reason, modern ferromagnetic MRAMs use more resistive spin valves with a tunnel barrier separating the free and the reference layer and showing $\sim$100\% TMRs.\cite{Parkin2003,Chappert2007} This has motivated studies of antiferromagnetic tunnel junctions. However, instead of the elusive non-relativistic antiferromagnetic TMR, experiments focused on the relativistic tunnelling AMR (TAMR).\cite{Shick2010,Park2011b,Wang2012a,Ralph2013,Petti2013,Wang2014e}  Unlike the antiferromagnetic GMR/TMR\cite{Nunez2006}, the TAMR devices can operate with only one magnetic electrode facing the tunnel barrier and hence do not rely on the subtle spin-coherent quantum-interference effects.\cite{Shick2010} 

\begin{figure}[h!]
\hspace*{-0cm}\epsfig{width=1\columnwidth,angle=0,file=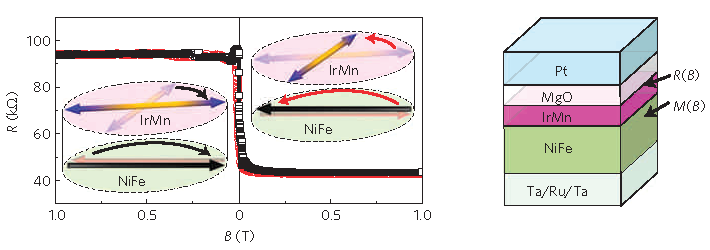}
\caption{{\bf a}, Larger than 100\% TAMR signal recorded in a NiFe/IrMn(1.5 nm)/MgO/Pt tunnel junction. The insets illustrate the rotation of antiferromagnetic moments in IrMn through the exchange-spring effect of the adjacent NiFe ferromagnet. {\bf b}, The external magnetic field is sensed by the NiFe ferromagnet whereas the tunnelling transport is governed by the IrMn antiferromagnet. From Ref.~\onlinecite{Park2011b}.}
\label{IrMn_TAMR}
\end{figure}

Bistable antiferromagnetic TAMR signals as large as 160\% have been reported at low temperatures\cite{Park2011b} and are illustrated in Fig~\ref{IrMn_TAMR}. The magnetic electrode in this antiferromagnetic TAMR device is formed by a conventional IrMn/NiFe bilayer, however, with an inverted order as compared to conventional TMR stacks. In the latter devices, the ferromagnet (NiFe) is placed in contact with the tunnel barrier and serves as the reference to the free ferromagnet on the other side of the barrier. The antiferromagnet (IrMn) is at the other interface of the reference ferromagnet forming a fixed exchage-bias structure. In the inverted structure used for the antiferromagnetic TAMR experiments, the IrMn antiferromagnet  is in contact with the barrier and by this governs the TAMR. Antiferromagnetic moments of IrMn are rotated via the exchange-spring effect from the NiFe ferromagnet at the opposite interface which is sensitive to weak magnetic fields (see Fig~\ref{IrMn_TAMR}).

Similar to the reading, efficient electrical writing schemes in antiferromagnets become feasible when introducing the relativistic spin-dependent phenomena into the antiferromagnetic spintronics.\cite{Zelezny2014} N\'eel's concept of local fields, extended to non-equilibrium properties of antiferromagnets,\cite{MacDonald2011}  remains central. However, relativistic quantum mechanics adds to it new robust means for controlling these local fields by global electrical currents without auxiliary reference ferromagnets in the structure and without relying on subtle quantum-interference effects at perfectly ordered interfaces of magnetic multilayers.\cite{Zelezny2014,Wadley2015} 

Compared to the angular momentum to angular momentum transfer governing the non-relativistic STT in magnetic multilayers, the spin-orbit coupling term in the Dirac equation allows for the additional {\em linear momentum to spin angular momentum transfer phenomena}. This opens the possibility of constructing spintronic devices with a single uniform magnetic component and with self-referencing schemes provided by the internal linear momentum to spin angular momentum transfer under applied electrical currents.\cite{Sinova2014} 

Experimental roots of this relativistic pillar of antiferromagnetic spintronics can be traced back to  one of the N\'eel's contemporaries, Clifford Shull. Apart from providing the first direct evidence of the antiferromagnetic order by neutron scattering\cite{Shull1949} (Nobel prize in 1994), he made earlier seminal experiments with electron beams.\cite{Shull1943} The experiments confirmed the validity of Dirac's relativistic quantum mechanics by observing Mott scattering of electrons from heavy nuclei.\cite{Mott1929} The counterpart of Mott scattering in condensed matter physics is the spin Hall effect (SHE).\cite{Sinova2014} It allows for turning  even a non-magnetic conductor into an efficient injector of spin current and for using it instead of the reference ferromagnetic polarizer in spin torque  devices.\cite{Miron2011b,Liu2012} 

As illustrated in Figs.~\ref{F-AF}d,e, a vertical spin current can be generated in a non-magnetic/antiferromagnetic stack due to the SHE in the spin-orbit coupled non-magnetic polarizer by an in-plane electrical current. The (anti)damping-like STT can efficiently reorient the antiferromagnet as in the case of the ferromagnetic polarizer and of injection by the vertical electrical current.\cite{Zelezny2014} Moreover, the SHE devices require no auxiliary ferromagnet in the structure and allow for a reversible electrical switching between the two orthogonal antiferromagnetic states by applying the writing electrical current along two orthogonal in-plane directions (see Figs.~\ref{F-AF}d,e). 

Still, the SHE stack geometry has some limitations. The torques are sensitive to the quality of the non-magnetic/antiferromagnetic interface and are efficient only for antiferromagnetic film thicknesses of the order of the spin diffusion length which in antiferromagnets is typically on the nanometer scale.\cite{Acharyya2011} Experimental indications of the presence of the SHE-induced torques have been recently reported at interfaces of ultra-thin IrMn with strongly spin-orbit coupled non-magnetic  Ta.\cite{Reichlova2015}

\begin{figure}[h!]
\hspace*{-0cm}\epsfig{width=1\columnwidth,angle=0,file=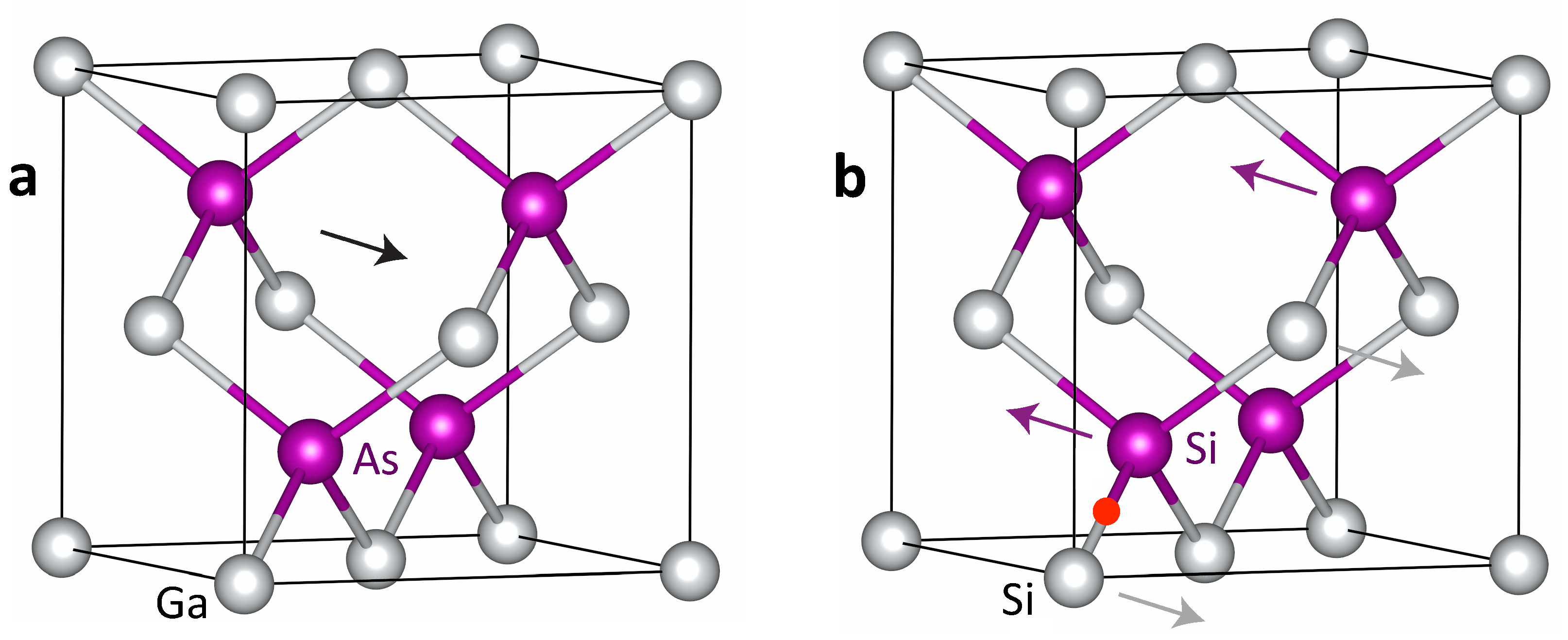}
\caption{{\bf a}, Global ferromagnetic-like non-equilibrium spin polarization generated by electrical current in a non-magnet lattice with global inversion-asymmetry (e.g. GaAs) due to the ISGE. {\bf b}, Local antiferromagnetic-like non-equilibrium spin polarization in a non-magnet lattice with local inversion-asymmetry (e.g. Si) due to the ISGE. Red dot shows the inversion-symmetry center of the Si lattice. The two Si atoms on either side of the center occupy inversion-partner lattice sites with locally asymmetric environments. In GaAs lattice, the inversion-symmetry center is absent since the two inversion-partner cites in the unit cell are occupied by different atoms }
\label{ISGE}
\end{figure}

The SHE was experimentally discovered a decade ago as a companion phenomenon to the inverse spin galvanic effect (ISGE). \cite{Kato2004d,Kato2004b,Wunderlich2004,Wunderlich2005}  The origin of the ISGE is also in the relativistic transfer between linear and spin angular momenta. Unlike the SHE generating a bulk spin-current and a resulting surface/interface spin polarization, ISGE induces a non-equilibrium spin polarization in the bulk of a crystal. It was experimentally discovered in GaAs\cite{Silov2004,Kato2004b,Ganichev2004b,Wunderlich2004,Wunderlich2005,Bernevig2005c,Chernyshov2009} where the non-equilibrium spin-polarization is globally uniform, as illustrated in Fig.~\ref{ISGE}. Apart from the spin-orbit coupling, the global current-induced spin polarizaton by ISGE requires a non-centrosymmetric unit cell of the crystal, as is the case of the zinc-blende GaAs. (For the cubic lattice of GaAs, an additional symmetry lowering is required by, e.g., a tetragnal deformation due to strain.) 

As in the zinc-blende lattices, the related diamond lattices of e.g. Si or Ge, shown in Fig.~\ref{ISGE}b, have two atoms in the unit cell  with locally non-centrosymmetric environments. The two atoms sitting on the inversion partner lattice sites are, however, identical which makes the diamond lattice unit cell globally centro-symmetric. As a result, the diamond lattice is an example where the ISGE can generate local non-equilibrium spin polarizations with opposite sign and equal magnitude on the two inversion-partner atoms while the global polarization integrated over the whole unit cell vanishes.\cite{Zelezny2014}  Here {\em a uniform electrical current induces a non-equilibrium antiferromagnetic spin polarization in the bulk crystal}. 

\begin{figure}[h!]
\hspace*{-0cm}\epsfig{width=1\columnwidth,angle=0,file=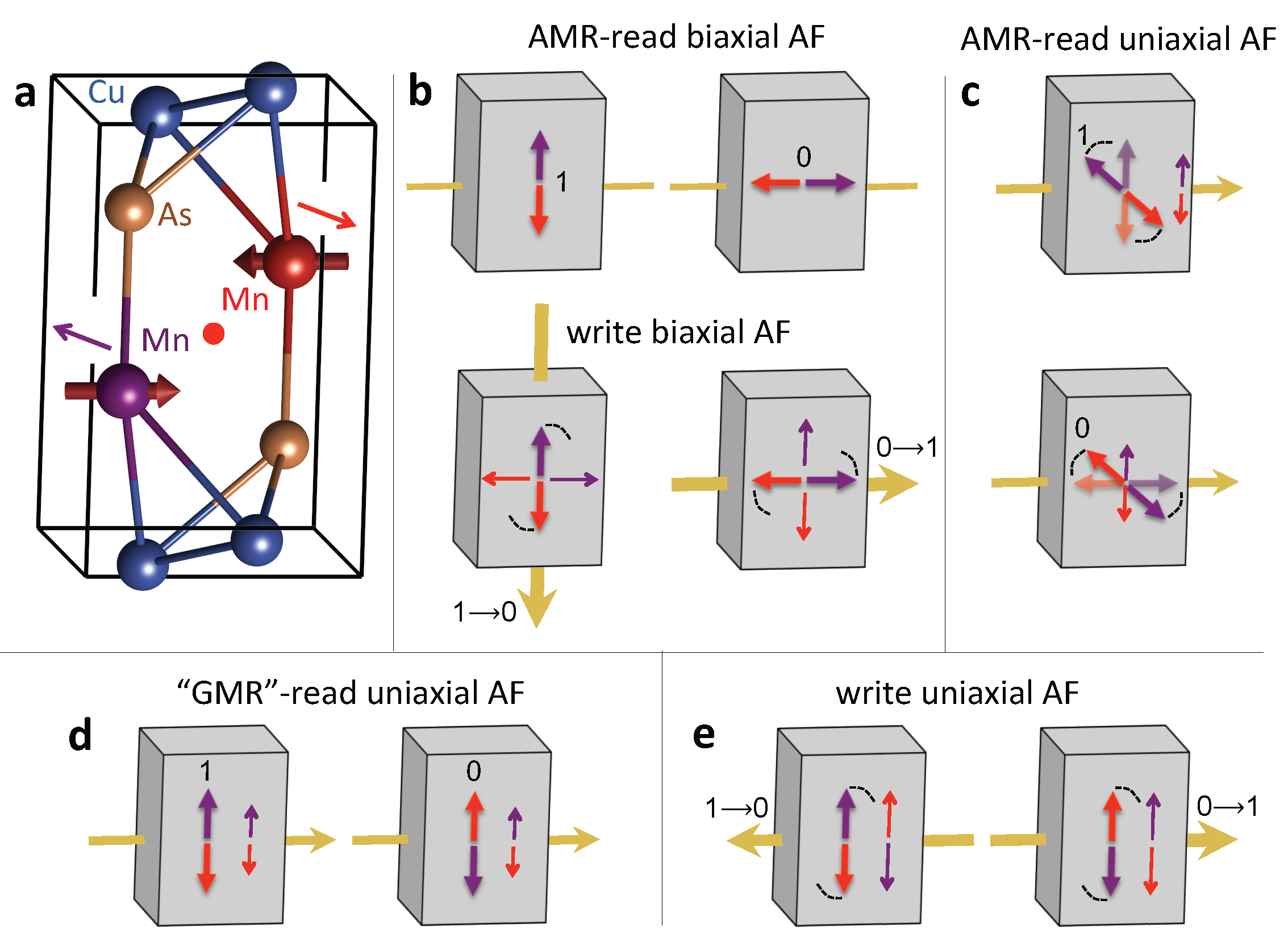}
\caption{{\bf a}, Local antiferromagnetic-like non-equilibrium spin-polarization inducing a local staggered effective field (thin arrows) in an antiferromagnet (thick arrows) lattice with local inversion-asymmetry (e.g. CuMnAs). {\bf b}, Top: electrical AMR reading of two stable orthogonal antiferromagnetic magnetizations representing 1 and 0 in a biaxial memory. Gold lines show probing current path. Bottom: Electrical writing of the states by a strong staggered effective field generated by one or the other orthogonal writing current paths (gold lines with arrows indicate current direction). {\bf c}, AMR reading of a uniaxial antiferromagnet assisted by a weaker current-induced staggered field that partially tilts the 1/0 state away/towards the current direction. {\bf d}, "GMR"-like reading of a uniaxial antiferromagnet assisted by a weaker current-induced staggered spin-polarization aligned/anti-aligned with sublattice-magnetizations in the 1/0 state. {\bf e}, Electrical writing of the uniaxial antiferromagnet by a strong current-induced staggered field by one or the other current polarity.}
\label{AF_MEM}
\end{figure}

In Si there is no equilibrium antiferromagnetic order that could be manipulated by these local staggered non-equilibrium polarizations. However, antiferromagnets like CuMnAs\cite{Wadley2013,Wadley2015} shown in Fig.~\ref{AF_MEM}a or Mn$_2$Au\cite{Shick2010,Wu2012,Barthem2013,Zelezny2014} share the crystal symmetry allowing for the current-induced staggered polarization whose sign alternates between the inversion-partner atoms. Moreover, one inversion-partner lattice site is occupied by the magnetic Mn belonging to the first antiferromagnetic spin sublattice and the other inversion partner to the second spin sublattice. Under the applied electrical current, a commensurate self-induced staggered polarization playing the role of the above STT's ${\bf p}_i$ is generated internally by the ISGE. As in the case of the non-relativistic STT, the field-like component of the relativistic spin torque induced by the ISGE can efficiently reorient the antiferromagnet for the case of staggered ${\bf p}_i$. 

Recent experiments have demonstrated\cite{Wadley2015} that the relativistic staggered fields can indeed couple as strongly to the N\'eel order as uniform fields couple to the global magnetic order in ferromagnets. It opens the possibility for constructing antiferromagnetic devices using analogous microelectronic designs to the ferromagnetic AMR-MRAMs\cite{Daughton1992} with the writing Oersted field replaced by the relativistic current-induced staggered field. 

Fig.~\ref{AF_MEM}b illustrates the AMR read-out combined with electrical writing by the staggered fields in a biaxial antiferromagnetic memory.\cite{Zelezny2014} Commercialized ferromagnetic AMR-MRAMs\cite{Daughton1992}  utilized uniaxial magnets with opposite magnetizations representing 1 and 0. Uniaxial magnets tend to have higher magnetic anisotropy barrier between the two memory states and, therefore, more robust storage than biaxial magnets. In this case the Oersted field was used also for reading by partially tilting the magnetization of one state towards and of the other state away from the reading current direction. This allowed to read the uniaxial memory states by the AMR. As illustrated in Fig.~\ref{AF_MEM}c, the current induced staggered fields allow for employing an analogous scheme in a uniaxial antiferromagnet.  Alternatively, the current-induced ISGE polarization can act as an internal self-reference for a GMR-like readout,\cite{Olejnik2015,Avci2015} without involving any tilt of the moments. At a larger applied current, a larger amplitude staggered ISGE-field can also allow for reversing the uniaxial antiferromagnet, as illustrated in Fig.~\ref{AF_MEM}d,e.

\subsection*{Generation, detection, and transmission of spin-currents in antiferromagnets}
Spin-current generation by the SHE mentioned in the previous section is not limited to non-magnetic materials.\cite{Sinova2014} It has been experimentally demonstrated that antiferromagnets with strong spin-orbit coupling can also act as efficient SHE spin injectors. Measurements showing electrical reorientation of a ferromagnet by the SHE induced in an adjacent antiferromagnet\cite{Tshitoyan2015,Fukami} highlight that antiferromagnetic spintronics aims not only at replacing ferromagnets but also at assisting ferromagnets in their performance in spintronic devices. We also note  that, according to recent theory predictions,\cite{Cheng2014c} even insulating antiferromagnets can act as efficient spin-current sources via the antiferromagnetic resonance spin pumping.

\begin{figure}[h!]
\hspace*{-0cm}\epsfig{width=1\columnwidth,angle=0,file=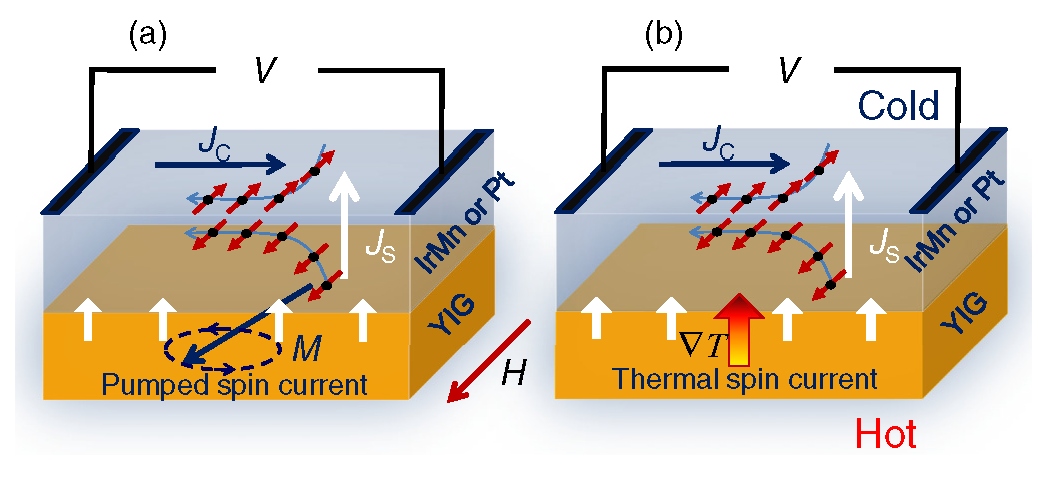}
\caption{Sketches showing the YIG/IrMn or YIG/Pt structures and the electrodes used to measure the dc voltage due to the ISHE  in IrMn (Pt) resulting from the spin currents generated in two configurations: {\bf a}, microwave FMR spin pumping from YIG and {\bf b}, longitudinal spin Seebeck effect. The static field $H$ is applied in the film plane. From Ref.~\onlinecite{Mendes2014}.}
\label{ISHE}
\end{figure}

Large inverse SHEs observed in transition metal antiferromagnets\cite{Mendes2014,Zhang2014e} imply their utility as spin-current detectors. Experiments illustrated in Fig.~\ref{ISHE} have demonstrated the detection by the inverse SHE in IrMn of spin currents generated either thermally by the spin Seebeck
effect or by microwave spin pumping from an adjacent insulating ferromagnet. 

Recent studies suggest that antiferromagnets can also act as efficient spin-current transmitters.\cite{Takei2014b,Wang2014d,Hahn2014}  The functionality has been studied primarily in insulating antiferromagnets where spin currents are transmitted by excitations of the local moments without involving charge transport. Theory predicts that antiferromagnets can support an essentially lossless superfluid spin transport over length-scales inverse proportional to the antiferromagnetic damping parameter.\cite{Takei2014b} 

\subsection*{Fast magnetic moment dynamics in antiferromagnets}
As already noted in the seminal N\'eel's works,\cite{Neel1970} reorientation of the antiferromagnetic moments involves canting of the two spin sublattices from their equilibrium antiparallel state which costs exchange energy.\cite{Keffer1952} In ferromagnets, a coherent reorientation of the magnetization involves no relative canting of the moments and the associated energy cost is only related to the much weaker magnetic anisotropy. As a result, antiferromagnets have typically much faster dynamics than ferromagnets.\cite{Keffer1952,Gomonay2014}

An illustration has been provided by optical experiments in insulating antiferromagnets.\cite{Kimel2004,Fiebig2008,Kampfrath2010,Satoh2010,Satoh2014} For example, picosecond-scale reorientation of the antiferromagnetic spin-axis was reported in an optical pump-and-probe study of  a rare-earth orthoferrite.\cite{Kimel2004} The origin of the generated staggered field was different than in current induced spin torques discussed above. The material has a  temperature dependent  antiferromagnetic easy-axis direction and the corresponding staggered anisotropy field was induced by laser-heating the sample above the easy-axis transition temperature. The microscopic origin of the staggered field is not crucial, however, for the time-scale of the spin-dynamics. The experiment therefore illustrates  that  the antiferromagnetic spin-axis reorientation in memory devices with electrical writing is not limited in principle by the antiferromagnetic spin dynamics itself but only by the circuitry time-scales for delivering electrical pulses which can reach $\sim100$~ps.\cite{Schumacher2003} 

Studies of the dynamics in the antiferromagnetic spintronic devices are not limited  to coherent reorientation of uniform antiferomagnetic domains. Recent theory works have  considered also schemes employing domain walls and other antiferromagnetic textures.\cite{Cheng2012a,Tveten2013,Cheng2014,Tveten2014}

\subsection*{Antiferromagnetic materials for spintronics}
The  research of insulating antiferromagnets is an example of the broad range of materials considered in the context of antiferromagnetic spintronics.  Simple transition metal oxides served as ideal model systems from the early days  of anifrromagnetism\cite{Neel1970} and led to the discovery of exchange bias.\cite{Nogues1999} Above we also mentioned the role of complex oxides in optical experiments in antiferromagnets. BiFeO$_3$ is another remarkable member of the family of insulating antiferromagnets. It combines a high-temperature magnetic order with ferroelectricity and offers a range of phenomena for spintronics stemming from the interplay of the two types of order in a multiferroic material.\cite{Sando2013} 

Antiferromagntic semiconductors are the natural candidates for integrating spintronics and traditional microelectronics functionalities in one material.\cite{Jungwirth2011,Cava2011} The synthesis of semiconductors with high-temperature ferromagnetic ordering of spins, which would simultaneously enable the conventional tunability of electronic properties and spintronic functionalities, remains a significant challenge.\cite{Dietl2014,Jungwirth2014}  On the other hand,  antiferromagnetic order occurs much more frequently  than ferromagnetic order, particularly in conjunction with semiconducting electronic structure,\cite{Maca2012} as illustrated in Tab.~\ref{tab1}. 

Recent studies have identified several candidate antiferromagnetic semiconductor materials, ranging from  counterparts of common zinc-blende or heusler compound semiconductors,\cite{Jungwirth2011,Wadley2013,Beleanu2013,Kriegner}  to perovskite semiconductor oxides.\cite{Kim2009a,Fina2014,Wang2014c}   A particular focus in this materials research has been on the preparation of thin epitaxial films and heterostructures as a prerequisite for the envisaged spintronic devices. Spintronic functionalities, including AMR read-out and storage, have been already demonstrated in several antiferromagnetic semiconductor structures.\cite{Fina2014,Wang2014c,Kriegner}

\begin{table}[h]
\begin{center}
\begin{tabular}{ccc|cccc}
\hline
II-VI & $T_c$~(K) & $T_N$~(K) &   III-V & $T_c$~(K) & $T_N$~(K) &\\
MnO & &122&FeN& &100 &\\
MnS& &152&FeP& &115&\\
MnSe& &173& FeAs& &77&\\
MnTe& &{\bf 323}& FeSb& &100-220&\\
EuO&67& &GdN&72& &\\
EuS&16& &GdP& &15&\\
EuSe& &5& GdAs& &19&\\
EuTe& &10& GdSb& &27&\\
\hline
I-VI-III-VI& & & II-V-IV-V & & &\\
CuFeO$_2$& &11&MnSiN$_2$& &{\bf 490}&\\
\cline{4-7}
CuFeS$_2$& &{\bf 825}& I-II-V& & &\\
CuFeSe$_2$& &70& LiMnAs& & {\bf 374}&\\
CuFeTe$_2$& &254& & & &\\

\hline
\end{tabular}
\label{table1}
\end{center}
\caption{Comparison of ferromagnetic Curie temperatures ($T_c$) and antiferromagnetic N\'eel temperatures ($T_N$) in II-VI, I-VI-III-VI, III-V, II-V-IV-V, and I-II-V magnetic semiconductors. From Refs.~\onlinecite{Maca2012,Beleanu2013} and references therein.}
\label{tab1}
\end{table}

In previous sections we already mentioned  examples of metal antiferromagnets which have so far driven much of the research in antiferromagnetic spintronics. Alloys of Ir and Mn are a prime example of metal antiferromagnets gradually progressing from favorable passive exchange-bias materials to active electrodes in TAMR devices,\cite{Park2011b,Wang2012a,Ralph2013,Wang2014e}  to  SHE injectors of spin current controlling adjacent ferromagnets,\cite{Tshitoyan2015,Fukami} or to sensitive spin detectors.\cite{Mendes2014}  

An additional remarkable property of IrMn$_3$ is that its non-collinear antiferromagnetic order with three compensated spin sublattices is expected to allow for the anomalous Hall effect (AHE).\cite{Chen2014} In ferromagnets,  AHE scales linearly with magnetization which suggests its absence in antiferromagnets. Its origin is described in terms of the broken time-reversal symmetry of the ordered state and spin-orbit coupling.\cite{Nagaosa2010} Antiferromagnets have also broken time reversal symmetry resulting in the sublattice magnetization. In collinear antiferromagnets, however, a time-reversal combined with translation recovers the symmetry and makes the AHE vanish. However, such a symmetry operation is not in general present in non-collinear antiferromagnets. In IrMn$_3$, the AHE is predicted to have comparable magnitude to ferromagnets.\cite{Chen2014}

To complete our brief excursion up the conductivity ladder we also recall that the antiferromagnetic order can coexist with superconductivity.\cite{Norman2011} Finally, the overview of antiferromagnetic materials would not be complete without mentioning synthetic antiferromagets.\cite{Bruno2002,Parkin2003} These man-made structures comprise ferromagnetic layers antiferromagnetically coupled through a metallic spacer. They led to the discovery of the GMR.\cite{Baibich1988,Binasch1989} 

In synthetic antiferromagnets, the interlayer coupling is typically orders of magnitude weaker than the exchange coupling between neighboring atoms in antiferromagnetic crystals. Several approaches known from crystal antiferromagnets have been successfully adopted in spintronic devices comprising synthetic antiferromagnets. For example, a spin-flop-like reorientation provided the basis for a reliable magnetic field writing of the free layer composed of a synthetic antiferromagnet in the commercial toggle MRAMs.\cite{Engel2005a} Reference layers prepared in the form of synthetic antiferromagnets, on the other hand, provided a better stability of these fixed magnets and suppressed the effects of stray field on the free magnetic electrode in the spin valve.\cite{Parkin2003}  In combination with relativistic spin-orbit coupling phenomena, synthetic antiferromagnets have recently led also to observations of a reliable and efficient current-driven domain-wall motion in racetrack memory devices.\cite{Saarikoski2014,Yang2015a} 

\subsection*{Concluding remarks}
Despite the recently rapidly growing literature on antiferromagnetic spintronics, the field is still at its infancy and it difficult to predict the future course of basic research  in the field and viable applications. Nevertheless, we can make a few remarks  based on the current knowledge and analogies with ferromagnetic spintronics.

Spin torques and magnetoresistances are now in principle available for electrical manipulation and detection of antiferromagnets. However, the area requires continuing research to reach a level of control allowing for fully exploiting the merits  of antiferromagnets in practical devices. For example, the high intrinsic frequencies of antiferromagnetic dynamics do not automatically guarantee ultra-fast switching. Antiferromagnetic domains play an essential role in this context which makes their detailed understanding an important challenge.\cite{Li2015a} Research in antiferromagnetic domain walls and other textures falls also naturally into this category of future studies.

The read-out speed and the size-scalability of spintronic devices is proportional to the magnitude of the magnetoresistance signal. Large TAMRs in antiferromagnetic tunnel junctions have so far been observed only at low temperatures. Increasing the temperature robustness of the effect is another challenge for future research. In ferromagnets, huge relativistic AMR-like effects have been observed in devices where the magnetic electrode is capacitively coupled to the transport channel which can be non-magnetic.\cite{Wunderlich2006,Ciccarelli2012} These types of devices with the gate electrode formed by a strongly spin-orbit coupled antiferromagnet will have a more complex design than ohmic or tunneling resistors but may provide a significant enhancement of the detection signal. Inverse transistor structures with a normal gate and the transport channel made of an aniferromagnetic semiconductor represent a complementary and yet another unexplored research direction.

The potential mid or long term applications of antiferromagnetic spintronics will depend on results of these and many other research directions in the field. One area of applied interest may be magnetic cloaking, i.e., making objects invisible to magnetic fields. Referring to real physical objects, this was already one of the many practical interests of N\'eel\cite{Coey2003} and it keeps intriguing scientists to date.\cite{Gomory2012} The new twist that antiferromagnetic spintronics  introduces to magnetic cloaking is making magnetically invisible not only the physical object, namely the magnetic medium or device, but also the information stored on the device. Here starting from a single bit may already initiate new paths to viable applications. Magnetic cards or security tags invisible to common scanners for ferromagnets and uneraseable by high magnetic fields are clearly foreseeable within the current knowledge in antiferromagnetic spintronics. 

For computer MRAMs, antiferromagnets may turn more practical than non-magnetic SHE layers in the lateral-current writing  schemes since their magnetic order  provides additional functionalities, such as the exchange-bias. Purely antiferromagnetic MRAMs with the writing and reading performances matching the most advanced ferromagnetic MRAMs are a challenge for a longer-term research. Still, with the functionalities available today, the antiferromagnetic counterpart of the early ferromagnetic AMR-MRAM is in principle feasible. It can be used to demonstrate the combination of the radiation-hardness,\cite{Daughton1992} genuine to  spin-based devices, with the exceptional magnetic-field-hardness of antiferromagnets. Finally, we anticipate that unique spin-transport characteristics of insulating antiferromagnets, including the predicted lossless spin-transmission, may find applications in spin interconnects and make antiferromagnets potentially attractive for  basic and applied research in magnonics.\cite{Serga2010}

%

\end{document}